\documentclass[conference]{IEEEtran}

\IEEEoverridecommandlockouts

\usepackage{cite}
\usepackage{amsmath,amssymb,amsfonts}
\usepackage{algorithm}
\usepackage{algpseudocode}
\usepackage{graphicx}
\usepackage{textcomp}
\usepackage{xcolor}
\usepackage{tikz}
\usetikzlibrary{positioning,decorations.pathreplacing}
\usepackage{multirow}
\usepackage{booktabs}
\usepackage{array}
\usepackage{url}
\usepackage{float}
\begin{document}

\title{FlowCLIP: Contrastive Pretraining Using Domain Names for Encrypted Traffic Classification}

\author{
\IEEEauthorblockA{
Eun Hun Choi\\
Department of Computer Science\\
University of North Carolina at Chapel Hill\\
Chapel Hill, NC, USA\\
paulchoi@cs.unc.edu}
}
\maketitle
\begin{abstract}
Network traffic classification enables website fingerprinting, intrusion detection, and quality of service management. However, developing methods that capture stable and generalizable traffic patterns remains challenging. We introduce FlowCLIP, a contrastive pretraining framework for learning traffic representations using only side-channel features: packet inter-arrival times, packet sizes, and packet directions. FlowCLIP feeds traffic flow features into a traffic encoder and the corresponding domain names into a text encoder, aligning their learned representations through a CLIP-style contrastive loss. After pretraining on the CESNET-QUIC22 dataset, we freeze the traffic encoder and evaluate it through linear probing. We also assess whether the pretrained representations transfer to a separate dataset, the UC Davis dataset. Through these evaluations, we show that domain names can be used directly for traffic representation learning without defining traffic classes.
\end{abstract}

\section{Introduction}
Network traffic classification supports a variety of tasks, including website fingerprinting, intrusion detection, and quality of service management \cite{website_fingerpriting,intrusion_detection,qos_survey_2021}. These tasks involve identifying websites visited by users, characterizing traffic associated with malicious activity, and allocating network resources efficiently across applications.

Two traditional network traffic classification methods are port-based identification, which is based on transport layer port numbers, and deep packet inspection, which reads packet content. In modern network environments, both methods have become unreliable: dynamic port negotiation limits identification by port numbers, while end-to-end encryption limits visible packet content \cite{qos_survey_2021}.

Recent advances in network infrastructure, transport security, and transport-layer protocols further limit the identifiers available to traffic classifiers, including server IP addresses, domain names, and transport-layer header information. Content delivery networks may host multiple services behind a single address, reducing the reliability of server IP addresses for application identification \cite{trevisan_does_2020}. Domain names remain reliable identifiers; however, DNS over HTTPS and TLS Encrypted Client Hello can hide these identifiers \cite{trevisan_does_2020,encrypted_hello}. In addition, major content providers have widely adopted QUIC, a UDP-based secure transport protocol that reveals less transport-layer header information to passive observers than TLS over TCP \cite{langley_quic_2017,mucke_waiting_2025,rfc9312}. Given these limitations, recent encrypted traffic classification studies increasingly train machine learning, deep learning, and self-supervised representation learning models using visible packet metadata and encrypted packet byte representations \cite{qos_survey_2021,wickramasinghe_less_2025,et-bert,reject_option,akbari_encrypted_traffic}. However, reported performance can be difficult to interpret for several reasons.

First, evaluation methodology and input representation can strongly influence the results. Packet-level splits may introduce leakage by allowing packets from the same flow to appear in both training and test sets, enabling models to rely on flow-specific artifacts rather than generalizable behavior \cite{zhao_sweet_2025,wickramasinghe_sok_2025}. Input representation poses a related challenge: when models are trained on packet payload bytes from datasets containing unencrypted traffic, they may learn payload-dependent signals that would not be available in encrypted traffic \cite{foundation_model_limit,wickramasinghe_sok_2025}. These issues motivate evaluations that split data at the flow-level and restrict inputs to features available under encryption.

Second, traffic classification commonly relies on manually defined application categories, but there is no standard procedure for mapping SNI domain names to these categories \cite{reject_option,akbari_encrypted_traffic,wickramasinghe_sok_2025,CESNETQUIC}. As a result, application labels may vary across datasets and reflect dataset-specific labeling choices \cite{labeling_bias}.

Third, recent studies show that lightweight machine learning models using flow-level features may perform competitively with more complex deep learning models on widely used traffic classification datasets \cite{pesek_lightweight_2025,wickramasinghe_less_2025,traffic_crisis}. For example, XGBoost \cite{xgboost_model} and 1-nearest neighbor classifiers (1-NN) \cite{1-nn} have been reported to match or exceed recent deep learning models on widely used datasets \cite{wickramasinghe_less_2025,pesek_lightweight_2025}. These findings suggest that new classification methods should be evaluated against strong baselines.

These challenges motivate a representation learning framework that uses features available under encryption, depends less on manually defined traffic classes when learning the traffic representations, and is evaluated against strong machine learning baselines.

Recent work has shown that a model pretrained using fine-grained domain classes can learn transferable traffic representations \cite{universalembedding}. We build on the insight of using domain names for traffic representation learning by designing a pretraining framework, FlowCLIP, that uses domain name as input alongside traffic flow features. FlowCLIP uses a text encoder and a traffic encoder to encode domain names and traffic flow features, respectively. FlowCLIP then applies a CLIP-style contrastive loss to align the traffic flow and domain name representations \cite{clip}. This approach enables the model to be pretrained without manually grouping domain names into fixed traffic classes.

To evaluate FlowCLIP in a realistic encrypted traffic setting, we use a large-scale QUIC traffic dataset because QUIC is widely deployed by major content providers and reveals less transport-layer information to passive observers than TCP \cite{langley_quic_2017,rfc9312,mucke_waiting_2025-1}. After pretraining, we freeze the traffic encoder and evaluate it through linear probing on canonicalized domain name labels. We compare FlowCLIP with competitive machine learning baselines--XGBoost and 1-NN--as well as a supervised model with the same architecture trained directly on the downstream classification task. All models are evaluated using a time-based protocol in which training data comes from earlier traffic and test data comes from later traffic \cite{traffic_crisis}. We also evaluate the pretrained encoder on a separate dataset to examine whether its representations transfer beyond the dataset used for pretraining. Therefore, this work studies whether the pretrained representations remain useful over time and across datasets.

The main contributions of this work are as follows:
\begin{itemize}
\item We present FlowCLIP, a contrastive pretraining framework that learns traffic flow representations by aligning side-channel features of encrypted traffic with domain name text through a CLIP-style objective.
\item We evaluate FlowCLIP under a realistic time-based protocol using only side-channel features and compare it with strong baselines, including XGBoost, 1-NN, and a supervised model with the same architecture. We also evaluate whether the pretrained traffic encoder transfers to a separate dataset.
\end{itemize}

\section{Background}
In this section, we provide background on the Domain Name System and the QUIC protocol.

\subsection{Domain Name System}
The Domain Name System provides a standard naming system for identifying hosts and other network resources using domain names \cite{domain_name}. Domain names are written from left to right as dot-separated labels, beginning with the most specific label and ending with the least specific label \cite{domain_name}.
\subsection{QUIC}
QUIC is a transport protocol that provides secure communication over UDP and has been widely adopted since 2021 by major service providers, including Akamai, Amazon, Apple, Cloudflare, Fastly, Google, Meta, and Microsoft \cite{mucke_waiting_2025}. QUIC reduces transmission latency by optimizing connection establishment and loss recovery \cite{langley_quic_2017}. Unlike TCP, which requires separate round trips for TCP and TLS handshakes, QUIC combines these procedures into a single round trip \cite{langley_quic_2017}. QUIC also mitigates the head-of-line blocking problem inherent in TCP, where packet loss delays the delivery of subsequent data until the missing packet is retransmitted, by enabling multiple independent streams within a single connection \cite{langley_quic_2017}. This isolates the effects of packet loss to the affected stream while allowing other streams to continue transmitting \cite{langley_quic_2017}.

\section{Challenges in Encrypted Network Traffic Classification}
In this section, we discuss three challenges that affect the development and evaluation of encrypted traffic classification methods: feature representation, evaluation setup, and application labeling.

\subsubsection*{Feature Representation}
Features used for traffic classification should reflect generalizable traffic patterns rather than dataset-specific artifacts \cite{dosanddont,wickramasinghe_sok_2025,emperor_has}. For example, models may rely on IP addresses, but these addresses can be tied to specific hosts, services, or content delivery networks at a particular time \cite{dosanddont,wickramasinghe_sok_2025,qos_appscanner_2016}. IP-based classifiers can fail when services change IP addresses over time, as observed for some CDN-hosted services \cite{ip-based}. The models that depend on dataset artifacts may perform well in controlled settings but degrade in real-world deployments \cite{emperor_has}.

A similar concern arises for raw packet byte representations. Although raw byte representations may reduce the need for manual feature engineering, they may also encode signals that do not generalize across encrypted sessions or deployment settings. Many packet header fields may contain session-specific information, and encrypted payload bytes vary across TLS sessions \cite{wickramasinghe_sok_2025,wickramasinghe_less_2025}. In addition, datasets with unencrypted payloads may encourage models to rely on features unavailable in encrypted traffic, while datasets with TLS metadata such as SNI may induce reliance on identifiers that may be unavailable when Encrypted Client Hello (ECH) is enabled \cite{foundation_model_limit,wickramasinghe_sok_2025}. These findings motivate the use of compact side-channel features that are observable under encryption and less tied to dataset-specific identifiers.

\subsubsection*{Evaluation Setup}
Some representation learning methods split traffic at the packet level, allowing packets from the same flow to appear in both the training and test sets \cite{zhao_sweet_2025}. Since these methods typically use raw packet bytes as features, the models trained using these methods may rely on flow-specific artifacts that are similar among packets from the same flow but differ across flows \cite{zhao_sweet_2025,wickramasinghe_sok_2025}. When evaluated with a flow-level split, where all packets from a given flow are assigned either to training or testing, the performance of the models drops substantially \cite{zhao_sweet_2025}. These findings highlight the importance of aligning the evaluation setup with the intended deployment setting, where models must generalize to entirely unseen traffic flows.

Furthermore, many widely used datasets are collected in controlled laboratory settings \cite{pesek_lightweight_2025}. As a result, even a random flow-level split may place nearly identical samples in both the training and test sets, making the classification task easier than it would be in deployment \cite{pesek_lightweight_2025}. This highlights the need to evaluate classification models on heterogeneous traffic that reflects realistic deployment conditions.

\subsubsection*{Application Labels}
Network traffic datasets require accurate ground truth labels because incorrect labels may introduce bias during training \cite{dosanddont}. Many encrypted traffic datasets use SNI domain names as a basis for assigning traffic labels \cite{reject_option,akbari_encrypted_traffic,wickramasinghe_sok_2025,CESNETQUIC}. However, defining application categories is challenging because there is no standard procedure for mapping domain names to applications. As a result, annotators may rely on different heuristics or levels of granularity.

For example, YouTube traffic may involve several related domains and subdomains, including \texttt{music.youtube.com}, \texttt{img.youtube.com}, and \texttt{googlevideo.com}. Although these domains are related to the YouTube platform, they do not necessarily correspond to the same service function. One annotator may group them under a broad label such as ``YouTube'' or ``Video Streaming,'' while another may assign finer-grained labels that distinguish music delivery, image hosting, and video content distribution. Since both approaches may be reasonable in the absence of a standard labeling procedure, application categories may differ across datasets. These challenges motivate pretraining the model backbone without relying on predefined labels and evaluating the learned representations using downstream labels defined separately for each dataset.
\section{Method}
In this section, we define the problem, describe the dataset and data preparation procedure, and present the FlowCLIP architecture and training objective.
\subsection{Problem Definition}
We analyze network traffic at the bidirectional flow level. A bidirectional flow consists of packets exchanged between two endpoints and is grouped using the forward and reverse 5-tuples (source IP address, destination IP address, source port, destination port, and transport protocol). Let \(\mathcal{X} = \{x_1, x_2, \dots, x_N\}\) denote a dataset of \(N\) flows, where each flow \(x_i\) is associated with a domain name \(s_i\). The objective of FlowCLIP is to learn a flow-level traffic encoder that produces traffic representations for downstream traffic classification tasks. During contrastive pretraining, FlowCLIP learns to align each traffic flow \(x_i\) with its corresponding domain name \(s_i\). We evaluate the learned traffic representations through domain name prediction using canonicalized domain name labels.
\subsection{Dataset}
We first discuss limitations of commonly used application classification datasets, then describe the dataset used in this study. Two popular datasets are ISCXVPN2016 and CSTNET-TLS1.3 \cite{iscx,et-bert}. ISCXVPN2016 contains traffic from six application types over regular and VPN connections, but it may not reflect modern encrypted traffic because it was released in 2016 and 98.9\% of its traffic is unencrypted \cite{wickramasinghe_sok_2025}. CSTNET-TLS1.3 contains TLS-encrypted TCP flows labeled with 120 website classes \cite{et-bert}. However, prior work notes that the SNI domains associated with individual flows are not available in the publicly released dataset \cite{zhao_sweet_2025}.

To evaluate FlowCLIP on heterogeneous network traffic, we use CESNET-QUIC22 dataset \cite{CESNETQUIC}. The dataset contains approximately 153 million QUIC flows collected over one month from an ISP backbone network, with each day of traffic provided as a separate CSV file. The dataset contains traffic from Weeks 44--47; for clarity, we refer to these periods as Weeks 1--4 throughout the work. For each flow, the dataset records sequential traffic features for up to 30 packets, including packet sizes, inter-arrival times, and directions. Packet sizes are reported in bytes, inter-arrival times are recorded in milliseconds, and directions are encoded as $+1$ for client-to-server traffic and $-1$ for server-to-client traffic. Each row also includes SNI domains, human-annotated application labels, and supplementary metadata.
\subsection{Data Preparation}
\subsubsection*{Feature Representation}
We use side-channel features: packet inter-arrival times, packet sizes, and packet directions \cite{akbari_encrypted_traffic}. Packet direction is encoded using signed packet size, where positive values denote client-to-server packets and negative values denote server-to-client packets. Each flow is represented using up to the first 30 packets, with shorter flows zero-padded, yielding a \(30 \times 2\) input sequence consisting of inter-arrival time and signed packet size.

To capture endpoint-specific behavior, we compute direction-aware inter-arrival time, defined as the elapsed time since the previous packet sent in the same direction. For the first packet in a flow, this value is set to 0. For the first packet from the opposite direction, its value is set to the inter-arrival time from the immediately preceding packet in the flow. Since the dataset reports inter-arrival time relative to the immediately preceding packet, we compute direction-aware inter-arrival time for subsequent same-direction packets by accumulating the inter-arrival times of intervening packets traveling in the opposite direction. This avoids measuring timing gaps between packets sent in opposite directions, which may belong to different QUIC streams.\footnote{A single QUIC connection may carry multiple concurrent streams.}

\subsubsection*{Domain Name Labels}
We use domain name labels for downstream classification task to reduce dependence on manually mapping traffic into application categories. However, domain name labels are long-tailed and may contain instance-specific identifiers that inflate the number of distinct labels \cite{trevisan_does_2020,universalembedding}. To construct a fixed label space, we automate canonicalization of the domain name using Week 1 data. Each domain name is decomposed into dot-separated components. For example, \texttt{mail.google.com} is decomposed into \texttt{mail}, \texttt{google}, and \texttt{com}. Components appearing at least twice in Week 1 are retained, while components appearing once are treated as out-of-vocabulary.\footnote{Repeated instance-specific components may therefore remain if they occur more than once in Week 1.} Each domain name is canonicalized by replacing out-of-vocabulary components with \texttt{<unk>}. We then retain the most frequent canonicalized domain names accounting for \(99\%\) of Week 1 traffic and map the remaining tail labels to an ``other'' class, yielding a total of 1,915 classes. For Weeks 2--4, we first canonicalize each domain name using the Week 1 component vocabulary and then map the resulting canonicalized domain name to one of the 1,915 fixed classes.

Week 1 data is split into training and validation sets using a stratified 90\%/10\% split over the canonicalized labels. This split is shared across contrastive pretraining, linear probing, and machine learning baseline training to keep the comparison controlled and evaluate whether the pretrained representation transfers to the downstream classification task. During contrastive pretraining, the selected flows are paired with their domain names, whereas linear probing uses the corresponding canonicalized labels. We use the same split across all stages for a controlled comparison, which means this is an experimental design rather than a requirement of the contrastive learning objective. In practice, FlowCLIP can be pretrained on a separate large-scale collection of flow--domain pairs with an independently defined validation split.
\subsection{FlowCLIP Architecture}
FlowCLIP consists of a traffic encoder and a domain name text encoder for contrastive pretraining. The traffic encoder is a Transformer encoder with self-attention and padding masks to ignore zero-padded positions \cite{transformer}. It takes side-channel features as input and mean-pools encoder outputs over non-padded positions to produce a flow-level representation.

The domain name text encoder represents each domain name as an ordered sequence of domain components. Components are ordered from top-level domain to subdomain; for example, \texttt{mail.google.com} is represented as [\texttt{com}, \texttt{google}, \texttt{mail}]. Each component is mapped to a learnable embedding vector shared across all occurrences of that component. A learnable positional embedding is added to each component embedding to encode its position in the domain hierarchy. The resulting component representations are mean-pooled over non-padded positions to produce a domain name representation.

\subsection{Training and Evaluation}\label{sec:training}
\subsubsection*{Contrastive Pretraining}
During pretraining, FlowCLIP aligns traffic flow embeddings with domain name embeddings. The traffic and domain name representations are first passed through separate projection layers of the respective encoders and then L2-normalized. Let \(z_i^{x}\) denote the resulting normalized traffic embedding for flow \(x_i\), and let \(z_i^{s}\) denote the resulting normalized domain name embedding for \(s_i\). Given a minibatch of traffic flows and their corresponding domain names, pairwise similarity logits are computed using a temperature-scaled dot product:
\[
\ell_{ij} = \frac{(z_i^{x})^\top z_j^{s}}{\tau},
\]
where \(\tau\) is the fixed temperature parameter, following the temperature scaling used in NT-Xent \cite{simclr}. The model is optimized by using the CLIP-style symmetric cross-entropy loss over the similarity logits \cite{clip}. This CLIP-style objective consists of two InfoNCE losses \cite{infonce}, where one contrasts each traffic flow representation against all domain name representations in the minibatch, while the other contrasts each domain name representation against all traffic flow representations. The two losses are averaged to obtain the overall symmetric loss.
\subsubsection*{Linear Probing}
After contrastive pretraining, the traffic encoder is frozen, the traffic encoder projection layer is replaced with a linear classification head, and only this head is trained. We perform linear probing by training the classification head with cross-entropy loss to predict canonicalized domain name labels. Since the encoder remains frozen, linear probing evaluates how well the pretrained traffic representations support the downstream classification task.
\section{Experiment}
In this section, we compare FlowCLIP with machine learning baselines, conduct ablation studies, and evaluate the cross-dataset transferability of the learned representations.
\subsection{Experimental Setup}
We use a time-based evaluation protocol inspired by \cite{traffic_crisis}. Week 1 is used for training and validation, and final evaluation is performed on Weeks 2--4. The train-validation split uses a fixed split seed of 42, and all models are trained using seed 0. We compare FlowCLIP against 1-NN and XGBoost. Table~\ref{tab:flowclip_hyperparameters} summarizes the FlowCLIP hyperparameters.
\begin{table}[H]
\centering
\caption{FlowCLIP hyperparameters.}
\label{tab:flowclip_hyperparameters}
\begin{tabular}{lcc}
\toprule
Category & Hyperparameter & Value \\
\midrule
\multirow{2}{*}{Architecture}
& \(d_{\mathrm{model}}\) / heads / layers & 256 / 4 / 4 \\
& FFN dimension / dropout & 1024 / 0.1 \\
\midrule
\multirow{3}{*}{Optimization}
& Batch size / learning rate & 1024 / \(3\times10^{-4}\) \\
& Weight decay & 0.01 \\
& Contrastive temperature & 0.07 \\
\midrule
\multirow{2}{*}{Training}
& Pretraining: epochs & 50\\
& Linear probing: epochs & 50\\
\bottomrule
\end{tabular}
\end{table}
For FlowCLIP, both contrastive pretraining and linear probing use early stopping with patience 10, and the checkpoint with the best validation loss is restored for evaluation. For both baselines, each fixed-length input sequence is first flattened into a one-dimensional vector. For 1-NN, we apply z-score normalization to each feature using the mean and standard deviation calculated from the Week 1 training data, since packet inter-arrival times and packet sizes are measured on different scales. 1-NN baseline uses one-nearest-neighbor search with Manhattan distance, following \cite{pesek_lightweight_2025}. For XGBoost, we use a multiclass softmax objective and keep the default hyperparameter setting, except for early stopping. We apply early stopping with a patience of 10 validation rounds and select the best boosting iteration based on validation performance.

We used PyTorch for Transformer-based models, FAISS for 1-NN, and the XGBoost library for XGBoost. Transformer-based and 1-NN experiments used one NVIDIA A6000 GPU and 8 CPU cores. XGBoost was run on 8 CPU cores.\footnote{XGBoost was run on CPU because GPU execution exceeded the available GPU memory in our experimental environment.}

\subsubsection*{Evaluation Metrics}
We report accuracy, weighted F1, and macro F1. Accuracy measures the overall fraction of correct predictions. Weighted F1 averages per-class F1 scores according to class frequency, while macro F1 averages per-class F1 scores uniformly across classes.
\subsection{Results}
Table~\ref{tab:week_results} reports performance on Weeks 2--4, corresponding to 42.6M, 33.8M, and 44.2M evaluation samples, respectively. FlowCLIP performs better than 1-NN and XGBoost across all three weeks on accuracy, weighted F1, and macro F1.
\begin{table}[h]
\centering
\caption{Performance comparison across Weeks 2--4. All metrics are reported as percentages.}
\label{tab:week_results}
\small
\setlength{\tabcolsep}{6pt}
\begin{tabular}{llccc}
\toprule
Week & Model & Acc. & W-F1 & M-F1 \\
\midrule
\multirow{3}{*}{Week 2}
& 1-NN & 71.17 & 71.42 & 27.56 \\
& XGBoost & 50.17 & 48.59 & 7.19 \\
& FlowCLIP & \textbf{81.38} & \textbf{82.98} & \textbf{44.91} \\
\midrule
\multirow{3}{*}{Week 3}
& 1-NN & 69.11 & 69.73 & 26.27 \\
& XGBoost & 42.81 & 41.78 & 6.59 \\
& FlowCLIP & \textbf{70.30} & \textbf{73.28} & \textbf{40.35} \\
\midrule
\multirow{3}{*}{Week 4}
& 1-NN & 68.19 & 68.81 & 25.14 \\
& XGBoost & 42.30 & 41.26 & 6.31 \\
& FlowCLIP & \textbf{69.55} & \textbf{72.54} & \textbf{39.15} \\
\bottomrule
\end{tabular}
\end{table}

Table~\ref{tab:runtime} reports training time and inference latency. For 1-NN, training time corresponds to FAISS index construction, while FlowCLIP training time includes both contrastive pretraining and linear probing. Under our experimental setup, FlowCLIP requires the highest training time. XGBoost has the lowest inference latency even when run on CPU, while 1-NN has the highest inference latency.

\begin{table}[h]
\centering
\caption{Runtime comparison. Training time is reported in seconds, and inference latency is reported in milliseconds per sample.}
\label{tab:runtime}
\small
\setlength{\tabcolsep}{5pt}
\begin{tabular}{llcc}
\toprule
Model & Week & Training Time & Inference Latency\\
\midrule
\multirow{3}{*}{1-NN}
& Week 2 & \multirow{3}{*}{0.30} & 1.002 \\
& Week 3 & & 1.000 \\
& Week 4 & & 1.000 \\
\midrule
\multirow{3}{*}{XGBoost}
& Week 2 & \multirow{3}{*}{9501.75} & 0.007 \\
& Week 3 & & 0.006 \\
& Week 4 & & 0.008 \\
\midrule
\multirow{3}{*}{FlowCLIP}
& Week 2 & \multirow{3}{*}{121127.13} & 0.030 \\
& Week 3 & & 0.030 \\
& Week 4 & & 0.030 \\
\bottomrule
\end{tabular}
\end{table}

\subsection{Ablation Study}
We conduct two ablation studies to examine the contribution of contrastive pretraining and the effect of the classification head. First, we compare FlowCLIP with a supervised Transformer classifier that uses the same traffic encoder architecture, mean-pooled flow representation, and linear classification head. The Transformer is trained end-to-end with cross-entropy loss directly on the canonicalized domain name labels. This architecture-controlled comparison evaluates whether contrastive pretraining with domain name supervision provides traffic representations that remain competitive with a model optimized directly for the downstream classification task. Second, we evaluate a FlowCLIP variant with a lightweight two-layer multilayer perceptron (MLP) classification head instead of a linear head. The MLP head has a hidden dimension of 256 and uses GELU activation with a dropout rate of 0.1 between the two linear layers. This comparison evaluates whether replacing the linear head with a lightweight MLP improves downstream classification task. We report the mean and standard deviation of performance across three random seeds: 0, 1, and 2.
\begin{table*}[h]
\centering
\caption{Ablation study on Weeks 2--4. All metrics are reported as percentages.}
\label{tab:ablation}
\small
\setlength{\tabcolsep}{5pt}
\begin{tabular}{llccc}
\toprule
Week & Model & Acc. & W-F1 & M-F1 \\
\midrule
\multirow{3}{*}{Week 2}
& Transformer & 80.95 $\pm$ 0.42 & 82.62 $\pm$ 0.24 & \textbf{51.02 $\pm$ 0.13} \\
& FlowCLIP + Linear Head & 81.18 $\pm$ 0.17 & 82.89 $\pm$ 0.14 & 44.85 $\pm$ 0.35 \\
& FlowCLIP + MLP Head & \textbf{82.25 $\pm$ 0.54} & \textbf{83.73 $\pm$ 0.31} & 49.70 $\pm$ 0.15 \\
\midrule
\multirow{3}{*}{Week 3}
& Transformer & 69.53 $\pm$ 1.20 & 72.80 $\pm$ 0.75 & \textbf{45.88 $\pm$ 0.13} \\
& FlowCLIP + Linear Head & 69.69 $\pm$ 0.53 & 72.91 $\pm$ 0.39 & 40.19 $\pm$ 0.30 \\
& FlowCLIP + MLP Head & \textbf{71.42 $\pm$ 1.50} & \textbf{74.31 $\pm$ 1.01} & 44.29 $\pm$ 0.21 \\
\midrule
\multirow{3}{*}{Week 4}
& Transformer & 68.56 $\pm$ 1.21 & 71.97 $\pm$ 0.71 & \textbf{44.32 $\pm$ 0.09} \\
& FlowCLIP + Linear Head & 68.90 $\pm$ 0.58 & 72.16 $\pm$ 0.35 & 38.97 $\pm$ 0.34 \\
& FlowCLIP + MLP Head & \textbf{70.53 $\pm$ 1.55} & \textbf{73.50 $\pm$ 1.02} & 42.96 $\pm$ 0.21 \\
\bottomrule
\end{tabular}
\end{table*}

Table~\ref{tab:ablation} shows that the Transformer achieves the highest mean macro F1 across all three weeks, indicating stronger performance on less frequent classes. FlowCLIP with a linear head achieves comparable mean accuracy and weighted F1 with the Transformer, but obtains lower macro F1. Using an MLP head improves mean accuracy and weighted F1 and brings macro F1 closer to that of the Transformer. These results indicate that contrastive pretraining produces a competitive traffic encoder, while increasing the capacity of the classification head enables it to better leverage the pretrained representations, particularly for less frequent classes.
\subsection{Cross-Dataset Transfer}
To examine whether FlowCLIP representations transfer beyond the main evaluation setting, we conduct an additional cross-dataset experiment. We compare the frozen FlowCLIP traffic encoder with a newly trained linear head against a Transformer with the same architecture trained from scratch on the UC Davis dataset using cross-entropy loss \cite{ucdavis}. The dataset contains traffic from five application classes: Docs, Drive, Music, Search, and YouTube. It is organized into a pretraining split and two retraining splits corresponding to human-triggered and script-triggered traffic. Following \cite{universalembedding}, we train on the pretraining split and report performance separately on the human-triggered and script-triggered retraining splits. 

We use the same experimental settings as in the main evaluation, with one exception: the maximum number of training epochs is increased to 2,500 while the early stopping patience is kept at 10, as linear probing requires more epochs to converge on this dataset. Since packet times in the UC Davis dataset are reported relative to the start of each flow, we compute direction-aware inter-arrival times from these relative timestamps and convert them to milliseconds. We run all experiments with seeds 0--2. For FlowCLIP, each run loads the pretrained encoder from the corresponding seed. We report the mean and standard deviation of performance across the three runs.
\begin{table*}[h]
\centering
\caption{Cross-dataset transfer results on the UC Davis dataset. All metrics are reported as percentages.}
\label{tab:ucdavis_transfer}
\small
\setlength{\tabcolsep}{5pt}
\begin{tabular}{llccc}
\toprule
Evaluation Set & Model & Acc. & W-F1 & M-F1 \\
\midrule
\multirow{2}{*}{Human-triggered}
& Transformer & 74.70 $\pm$ 1.20 & 71.20 $\pm$ 1.56 & 70.61 $\pm$ 1.79 \\
& FlowCLIP + Linear Head & \textbf{82.33 $\pm$ 3.68} & \textbf{81.92 $\pm$ 4.15} & \textbf{82.09 $\pm$ 4.13} \\
\midrule
\multirow{2}{*}{Script-triggered}
& Transformer & \textbf{94.89 $\pm$ 0.77} & \textbf{94.87 $\pm$ 0.78} & \textbf{94.87 $\pm$ 0.78} \\
& FlowCLIP + Linear Head & 92.00 $\pm$ 0.67 & 91.89 $\pm$ 0.81 & 91.89 $\pm$ 0.81 \\
\bottomrule
\end{tabular}
\end{table*}

Table~\ref{tab:ucdavis_transfer} reports cross-dataset transfer results on the UC Davis dataset. FlowCLIP performs better than the Transformer on the human-triggered traffic across all metrics, suggesting that the pretrained encoder provides better generalization to human-triggered traffic in this setting. In contrast, the Transformer achieves higher performance on the script-triggered traffic.
\section{Discussion}
The experimental results provide several insights into both the strengths and limitations of FlowCLIP. Across all three models, performance decreases from Week 2 to later weeks, highlighting the impact of temporal distribution shift. Within each week, macro F1 also remains substantially lower than accuracy and weighted F1 across all models, suggesting that rare classes remain difficult to classify regardless of the method.

\subsection{Machine Learning Baselines} XGBoost had lower performance than FlowCLIP and 1-NN in our evaluation. Early stopping selected the first boosting round as the best iteration, despite a patience of 10 rounds. One possible explanation is the class imbalance introduced by the domain name canonicalization procedure, which produced domain classes with very few samples. This imbalance may have caused XGBoost to favor the majority classes and limited the benefit of additional boosting rounds. In contrast, 1-NN remained competitive with FlowCLIP in Weeks 3 and 4 despite its simplicity, but had higher inference latency than both FlowCLIP and XGBoost.
\subsection{Cross-Dataset Transfer}
The cross-dataset transfer results indicate that the effectiveness of FlowCLIP's pretrained encoder depends on the evaluation setting. FlowCLIP outperforms a Transformer trained from scratch on human-triggered traffic, but the Transformer performs better on script-triggered traffic. One possible explanation is that FlowCLIP is pretrained on traffic collected from an ISP network, which may more closely resemble human-triggered traffic than scripted-triggered traffic.
\subsection{Extensions}
Since the traffic flow embeddings and domain name embeddings share the representation space, FlowCLIP may support retrieval-based classification. A new traffic flow embedding can be compared with the text embeddings of candidate domain names using cosine similarity, and the domain name with the highest similarity can be selected as the prediction.
\subsection{Limitations}
This study has several limitations. First, the automated canonicalization procedure used to define the fixed label space reduces dependence on manually defined application categories, but it does not merge all instance-specific domain components. Stricter canonicalization could incorporate human-defined rules to further merge instance-specific identifiers such as shard labels similar to \cite{universalembedding}, but such extensions may introduce a tradeoff between reducing the long tail of labels and increasing dependence on manual design choices. Second, the domain name encoder used in FlowCLIP is deliberately lightweight, representing each domain as an ordered sequence of dot-separated components to keep the pretraining design simple. This captures the hierarchical structure of domain names, but it does not capture finer patterns within individual components. Such patterns may carry additional semantic information and may also be relevant to malicious domain detection \cite{dga}. For example, hyphenated components in domains such as \texttt{s-pinimg-com-cdn-cloudflare-net.pinimg.com} may encode information about the CDN infrastructure used by the service, while domains such as \texttt{kahoot-com.translate.goog} may encode an original hostname inside a single component. Since the domain name encoder is modular, future work could use hybrid text encoders that process domain names at both the domain component and character levels to better capture such patterns and improve classification. Third, FlowCLIP improves performance over the evaluated baselines, but performance still decreases on later weeks, indicating that contrastive pretraining followed by linear probing does not entirely address changes in traffic patterns over time. Future work could incorporate FlowCLIP into continual learning frameworks to handle temporal drift.

\section{Conclusion}
We introduced FlowCLIP, a pretraining framework that uses a CLIP-style contrastive loss to learn traffic representations directly from domain names available during training. We evaluate the pretrained representations under time-based evaluation and across datasets. This approach points towards possibility in scalable pretraining and reusable traffic encoders for classification tasks involving domain name-based labels. However, FlowCLIP's performance degradation over time indicates that temporal distribution drift remains an important challenge.
\bibliographystyle{IEEEtran}
\bibliography{sample-base}

@String{Computing = "Computing" }

@String{Computer = "{IEEE} Computer" }

@misc{simclr,
      title={A Simple Framework for Contrastive Learning of Visual Representations}, 
      author={Ting Chen and Simon Kornblith and Mohammad Norouzi and Geoffrey Hinton},
      year={2020},
      eprint={2002.05709},
      archivePrefix={arXiv},
      primaryClass={cs.LG},
      url={https://arxiv.org/abs/2002.05709}, 
}

@misc{infonce,
      title={Representation Learning with Contrastive Predictive Coding}, 
      author={Aaron van den Oord and Yazhe Li and Oriol Vinyals},
      year={2019},
      eprint={1807.03748},
      archivePrefix={arXiv},
      primaryClass={cs.LG},
      url={https://arxiv.org/abs/1807.03748}, 
}

@INPROCEEDINGS{ucdavis,
  author={Tong, Van and Tran, Hai Anh and Souihi, Sami and Mellouk, Abdelhamid},
  booktitle={2018 IEEE Global Communications Conference (GLOBECOM)}, 
  title={A Novel QUIC Traffic Classifier Based on Convolutional Neural Networks}, 
  year={2018},
  volume={},
  number={},
  pages={1-6},
  doi={10.1109/GLOCOM.2018.8647128}}

@INPROCEEDINGS{ip-based,
  author={Luxemburk, Jan and Hynek, Karel and Čejka, Tomáš},
  booktitle={2023 7th Network Traffic Measurement and Analysis Conference (TMA)}, 
  title={Encrypted traffic classification: the QUIC case}, 
  year={2023},
  volume={},
  number={},
  pages={1-10},
  doi={10.23919/TMA58422.2023.10199052}}

@misc{dga,
      title={Predicting Domain Generation Algorithms with Long Short-Term Memory Networks}, 
      author={Jonathan Woodbridge and Hyrum S. Anderson and Anjum Ahuja and Daniel Grant},
      year={2016},
      eprint={1611.00791},
      archivePrefix={arXiv},
      primaryClass={cs.CR},
      url={https://arxiv.org/abs/1611.00791}, 
}

@InProceedings{clip,
  title = 	 {Learning Transferable Visual Models From Natural Language Supervision},
  author =       {Radford, Alec and Kim, Jong Wook and Hallacy, Chris and Ramesh, Aditya and Goh, Gabriel and Agarwal, Sandhini and Sastry, Girish and Askell, Amanda and Mishkin, Pamela and Clark, Jack and Krueger, Gretchen and Sutskever, Ilya},
  booktitle = 	 {Proceedings of the 38th International Conference on Machine Learning},
  pages = 	 {8748--8763},
  year = 	 {2021},
  editor = 	 {Meila, Marina and Zhang, Tong},
  volume = 	 {139},
  series = 	 {Proceedings of Machine Learning Research},
  month = 	 {18--24 Jul},
  publisher =    {PMLR},
  url = 	 {https://proceedings.mlr.press/v139/radford21a.html}
}

@misc{rfc9312,
    series =    {Request for Comments},
    number =    9312,
    howpublished =  {RFC 9312},
    publisher = {RFC Editor},
    doi =       {10.17487/RFC9312},
    url =       {https://www.rfc-editor.org/info/rfc9312},
    author =    {Mirja Kühlewind and Brian Trammell},
    title =     {{Manageability of the QUIC Transport Protocol}},
    pagetotal = 29,
    year =      2022,
    month =     sep,
}

@inproceedings{zhao_sweet_2025,
	address = {New York, NY, USA},
	series = {{SIGCOMM} '25},
	title = {The {Sweet} {Danger} of {Sugar}: {Debunking} {Representation} {Learning} for {Encrypted} {Traffic} {Classification}},
	isbn = {979-8-4007-1524-2},
	shorttitle = {The {Sweet} {Danger} of {Sugar}},
	url = {https://dl.acm.org/doi/10.1145/3718958.3750498},
	doi = {10.1145/3718958.3750498},
	urldate = {2026-01-15},
	booktitle = {Proceedings of the {ACM} {SIGCOMM} 2025 {Conference}},
	publisher = {Association for Computing Machinery},
	author = {Zhao, Yuqi and Dettori, Giovanni and Boffa, Matteo and Vassio, Luca and Mellia, Marco},
	month = aug,
	year = {2025},
	pages = {296--310},
}

@inproceedings{wickramasinghe_less_2025,
	address = {New York, NY, USA},
	series = {{WWW} '25},
	title = {Less is {More}: {Simplifying} {Network} {Traffic} {Classification} {Leveraging} {RFCs}},
	isbn = {979-8-4007-1331-6},
	shorttitle = {Less is {More}},
	url = {https://dl.acm.org/doi/10.1145/3701716.3715492},
	doi = {10.1145/3701716.3715492},
	urldate = {2026-01-15},
	booktitle = {Companion {Proceedings} of the {ACM} on {Web} {Conference} 2025},
	publisher = {Association for Computing Machinery},
	author = {Wickramasinghe, Nimesha and Shaghaghi, Arash and Ferrari, Elena and Jha, Sanjay},
	month = may,
	year = {2025},
	pages = {1398--1401},
}

@inproceedings{wickramasinghe_sok_2025,
	title = {{SoK}: {Decoding} the {Enigma} of {Encrypted} {Network} {Traffic} {Classifiers}},
	issn = {2375-1207},
	shorttitle = {{SoK}},
	url = {https://ieeexplore.ieee.org/abstract/document/11023502},
	doi = {10.1109/SP61157.2025.00165},
	urldate = {2026-01-15},
	booktitle = {2025 {IEEE} {Symposium} on {Security} and {Privacy} ({SP})},
	author = {Wickramasinghe, Nimesha and Shaghaghi, Arash and Tsudik, Gene and Jha, Sanjay},
	month = may,
	year = {2025},
	pages = {1825--1843},
}

@inproceedings{pesek_lightweight_2025,
	title = {Lightweight {Traffic} {Classification}: {A} {Simple} {Baseline} {Matching} {Deep} {Learning} {Performance}},
	shorttitle = {Lightweight {Traffic} {Classification}},
	url = {https://ieeexplore.ieee.org/document/11096965},
	doi = {10.23919/TMA66427.2025.11096965},
	urldate = {2026-01-15},
	booktitle = {2025 9th {Network} {Traffic} {Measurement} and {Analysis} {Conference} ({TMA})},
	author = {Pesek, Jaroslav and Luxemburk, Jan and Hynek, Karel},
	month = jun,
	year = {2025},
	pages = {1--4},
}

@article{mucke_waiting_2025,
	title = {Waiting for {QUIC}: {Passive} {Measurements} to {Understand} {QUIC} {Deployments}},
	volume = {3},
	shorttitle = {Waiting for {QUIC}},
	url = {https://dl.acm.org/doi/10.1145/3768988},
	doi = {10.1145/3768988},
	number = {CoNEXT4},
	urldate = {2026-02-03},
	journal = {Proc. ACM Netw.},
	author = {Mücke, Jonas and Nawrocki, Marcin and Hiesgen, Raphael and Sattler, Patrick and Zirngibl, Johannes and Carle, Georg and Luxemburk, Jan and Schmidt, Thomas C. and Wählisch, Matthias},
	month = nov,
	year = {2025},
	pages = {41:1--41:26},
}

@article{CESNETQUIC,
	title = {{CESNET}-{QUIC22}: {A} large one-month {QUIC} network traffic dataset from backbone lines},
	volume = {46},
	issn = {23523409},
	shorttitle = {{CESNET}-{QUIC22}},
	url = {https://linkinghub.elsevier.com/retrieve/pii/S2352340923000069},
	doi = {10.1016/j.dib.2023.108888},
	language = {en},
	urldate = {2026-02-03},
	journal = {Data in Brief},
	author = {Luxemburk, Jan and Hynek, Karel and Čejka, Tomáš and Lukačovič, Andrej and Šiška, Pavel},
	month = feb,
	year = {2023},
	pages = {108888},
}

@article{trevisan_does_2020,
	title = {Does domain name encryption increase users' privacy?},
	volume = {50},
	issn = {0146-4833},
	url = {https://dl.acm.org/doi/10.1145/3411740.3411743},
	doi = {10.1145/3411740.3411743},
	language = {en},
	number = {3},
	urldate = {2026-02-10},
	journal = {ACM SIGCOMM Computer Communication Review},
	author = {Trevisan, Martino and Soro, Francesca and Mellia, Marco and Drago, Idilio and Morla, Ricardo},
	month = jul,
	year = {2020},
	pages = {16--22},
}

@article{mucke_waiting_2025-1,
author = {M\"{u}cke, Jonas and Nawrocki, Marcin and Hiesgen, Raphael and Sattler, Patrick and Zirngibl, Johannes and Carle, Georg and Luxemburk, Jan and Schmidt, Thomas C. and W\"{a}hlisch, Matthias},
title = {Waiting for QUIC: Passive Measurements to Understand QUIC Deployments},
year = {2025},
issue_date = {December 2025},
publisher = {Association for Computing Machinery},
address = {New York, NY, USA},
volume = {3},
number = {CoNEXT4},
url = {https://doi.org/10.1145/3768988},
doi = {10.1145/3768988},
journal = {Proc. ACM Netw.},
month = nov,
articleno = {41},
numpages = {26}
}

@inproceedings{langley_quic_2017,
	address = {Los Angeles CA USA},
	title = {The {QUIC} {Transport} {Protocol}: {Design} and {Internet}-{Scale} {Deployment}},
	isbn = {978-1-4503-4653-5},
	shorttitle = {The {QUIC} {Transport} {Protocol}},
	url = {https://dl.acm.org/doi/10.1145/3098822.3098842},
	doi = {10.1145/3098822.3098842},
	language = {en},
	urldate = {2026-02-12},
	booktitle = {Proceedings of the {Conference} of the {ACM} {Special} {Interest} {Group} on {Data} {Communication}},
	publisher = {ACM},
	author = {Langley, Adam and Riddoch, Alistair and Wilk, Alyssa and Vicente, Antonio and Krasic, Charles and Zhang, Dan and Yang, Fan and Kouranov, Fedor and Swett, Ian and Iyengar, Janardhan and Bailey, Jeff and Dorfman, Jeremy and Roskind, Jim and Kulik, Joanna and Westin, Patrik and Tenneti, Raman and Shade, Robbie and Hamilton, Ryan and Vasiliev, Victor and Chang, Wan-Teh and Shi, Zhongyi},
	month = aug,
	year = {2017},
	pages = {183--196},
}

@inproceedings{qos_appscanner_2016,
	address = {Saarbrucken},
	title = {{AppScanner}: {Automatic} {Fingerprinting} of {Smartphone} {Apps} from {Encrypted} {Network} {Traffic}},
	isbn = {978-1-5090-1751-5 978-1-5090-1752-2},
	shorttitle = {{AppScanner}},
	url = {http://ieeexplore.ieee.org/document/7467370/},
	doi = {10.1109/EuroSP.2016.40},
	language = {en},
	urldate = {2026-02-23},
	booktitle = {2016 {IEEE} {European} {Symposium} on {Security} and {Privacy} ({EuroS}\&{P})},
	publisher = {IEEE},
	author = {Taylor, Vincent F. and Spolaor, Riccardo and Conti, Mauro and Martinovic, Ivan},
	month = mar,
	year = {2016},
	pages = {439--454},
}

@article{qos_survey_2021,
	title = {A {Survey} on {Encrypted} {Network} {Traffic} {Analysis} {Applications}, {Techniques}, and {Countermeasures}},
	volume = {54},
	issn = {0360-0300},
	url = {https://dl.acm.org/doi/10.1145/3457904},
	doi = {10.1145/3457904},
	number = {6},
	urldate = {2026-02-23},
	journal = {ACM Comput. Surv.},
	author = {Papadogiannaki, Eva and Ioannidis, Sotiris},
	month = jul,
	year = {2021},
	pages = {123:1--123:35},
}

@inproceedings{et-bert, series={WWW ’22},
   title={ET-BERT: A Contextualized Datagram Representation with Pre-training Transformers for Encrypted Traffic Classification},
   url={http://dx.doi.org/10.1145/3485447.3512217},
   DOI={10.1145/3485447.3512217},
   booktitle={Proceedings of the ACM Web Conference 2022},
   publisher={ACM},
   author={Lin, Xinjie and Xiong, Gang and Gou, Gaopeng and Li, Zhen and Shi, Junzheng and Yu, Jing},
   year={2022},
   month=apr, pages={633–642},
   collection={WWW ’22} }

@article{akbari_encrypted_traffic,
author = {Akbari, Iman and Salahuddin, Mohammad A. and Aniva, Leni and Limam, Noura and Boutaba, Raouf and Mathieu, Bertrand and Moteau, Stephanie and Tuffin, Stephane},
title = {Traffic classification in an increasingly encrypted web},
year = {2022},
issue_date = {October 2022},
publisher = {Association for Computing Machinery},
address = {New York, NY, USA},
volume = {65},
number = {10},
issn = {0001-0782},
url = {https://doi.org/10.1145/3559439},
doi = {10.1145/3559439},
journal = {Commun. ACM},
month = sep,
pages = {75–83},
numpages = {9}
}

@article{reject_option,
   title={Fine-grained TLS services classification with reject option},
   volume={220},
   ISSN={1389-1286},
   url={http://dx.doi.org/10.1016/j.comnet.2022.109467},
   DOI={10.1016/j.comnet.2022.109467},
   journal={Computer Networks},
   publisher={Elsevier BV},
   author={Luxemburk, Jan and Čejka, Tomáš},
   year={2023},
   month=jan, pages={109467} }

@article{foundation_model_limit,
 title={Demystifying Network Foundation Models},
 author={Beltiukov, Roman and Guthula, Satyandra and Guo, Wenbo and Willinger, Walter and Gupta, Arpit},
 journal={Advances in neural information processing systems (NeurIPS)},
 year={2025}
}

@inproceedings{iscx,
author = {Habibi Lashkari, Arash and Draper Gil, Gerard and Mamun, Mohammad and Ghorbani, Ali},
year = {2016},
month = {02},
pages = {},
title = {Characterization of Encrypted and VPN Traffic Using Time-Related Features},
doi = {10.5220/0005740704070414}
}

@misc{traffic_crisis,
      title={When Simple Model Just Works: Is Network Traffic Classification in Crisis?}, 
      author={Kamil Jerabek and Jan Luxemburk and Richard Plny and Josef Koumar and Jaroslav Pesek and Karel Hynek},
      year={2025},
      eprint={2506.08655},
      archivePrefix={arXiv},
      primaryClass={cs.LG},
      url={https://arxiv.org/abs/2506.08655}, 
}

@misc{domain_name,
    series =    {Request for Comments},
    number =    1034,
    howpublished =  {RFC 1034},
    publisher = {RFC Editor},
    doi =       {10.17487/RFC1034},
    url =       {https://www.rfc-editor.org/info/rfc1034},
    author =    {Paul Mockapetris},
    title =     {{Domain Names - Concepts and Facilities}},
    pagetotal = 55,
    year =      1987,
    month =     nov,
}

@inproceedings{transformer,
	title = {Attention is {All} you {Need}},
	volume = {30},
	url = {https://proceedings.neurips.cc/paper_files/paper/2017/file/3f5ee243547dee91fbd053c1c4a845aa-Paper.pdf},
	booktitle = {Advances in {Neural} {Information} {Processing} {Systems}},
	publisher = {Curran Associates, Inc.},
	author = {Vaswani, Ashish and Shazeer, Noam and Parmar, Niki and Uszkoreit, Jakob and Jones, Llion and Gomez, Aidan N and Kaiser, Lukasz and Polosukhin, Illia},
	editor = {Guyon, I. and Luxburg, U. Von and Bengio, S. and Wallach, H. and Fergus, R. and Vishwanathan, S. and Garnett, R.},
	year = {2017},
}

@inproceedings{labeling_bias,
	address = {Hong Kong, China},
	title = {Are {We} {Modeling} the {Task} or the {Annotator}? {An} {Investigation} of {Annotator} {Bias} in {Natural} {Language} {Understanding} {Datasets}},
	shorttitle = {Are {We} {Modeling} the {Task} or the {Annotator}?},
	url = {https://aclanthology.org/D19-1107/},
	doi = {10.18653/v1/D19-1107},
	urldate = {2026-03-10},
	booktitle = {Proceedings of the 2019 {Conference} on {Empirical} {Methods} in {Natural} {Language} {Processing} and the 9th {International} {Joint} {Conference} on {Natural} {Language} {Processing} ({EMNLP}-{IJCNLP})},
	publisher = {Association for Computational Linguistics},
	author = {Geva, Mor and Goldberg, Yoav and Berant, Jonathan},
	editor = {Inui, Kentaro and Jiang, Jing and Ng, Vincent and Wan, Xiaojun},
	month = nov,
	year = {2019},
	pages = {1161--1166},
}

@article{universalembedding,
   title={Universal Embedding Function for Traffic Classification via QUIC Domain Recognition Pretraining: A Transfer Learning Success},
   volume={23},
   ISSN={2373-7379},
   url={http://dx.doi.org/10.1109/TNSM.2025.3642984},
   DOI={10.1109/tnsm.2025.3642984},
   journal={IEEE Transactions on Network and Service Management},
   publisher={Institute of Electrical and Electronics Engineers (IEEE)},
   author={Luxemburk, Jan and Hynek, Karel and Plný, Richard and Čejka, Tomáš},
   year={2026},
   pages={1647–1663} }

@inproceedings {dosanddont,
author = {Daniel Arp and Erwin Quiring and Feargus Pendlebury and Alexander Warnecke and Fabio Pierazzi and Christian Wressnegger and Lorenzo Cavallaro and Konrad Rieck},
title = {Dos and Don{\textquoteright}ts of Machine Learning in Computer Security},
booktitle = {31st USENIX Security Symposium (USENIX Security 22)},
year = {2022},
isbn = {978-1-939133-31-1},
address = {Boston, MA},
pages = {3971--3988},
url = {https://www.usenix.org/conference/usenixsecurity22/presentation/arp},
publisher = {USENIX Association},
month = aug
}

@misc{encrypted_hello,
    series =    {Request for Comments},
    number =    9849,
    howpublished =  {RFC 9849},
    publisher = {RFC Editor},
    doi =       {10.17487/RFC9849},
    url =       {https://www.rfc-editor.org/info/rfc9849},
    author =    {Eric Rescorla and Kazuho Oku and Nick Sullivan and Christopher A. Wood},
    title =     {{TLS Encrypted Client Hello}},
    pagetotal = 44,
    year =      2026,
    month =     mar,
}

@inproceedings{emperor_has,
author = {Jacobs, Arthur S. and Beltiukov, Roman and Willinger, Walter and Ferreira, Ronaldo A. and Gupta, Arpit and Granville, Lisandro Z.},
title = {AI/ML for Network Security: The Emperor has no Clothes},
year = {2022},
isbn = {9781450394505},
publisher = {Association for Computing Machinery},
address = {New York, NY, USA},
url = {https://doi.org/10.1145/3548606.3560609},
doi = {10.1145/3548606.3560609},
booktitle = {Proceedings of the 2022 ACM SIGSAC Conference on Computer and Communications Security},
pages = {1537–1551},
numpages = {15},
location = {Los Angeles, CA, USA},
series = {CCS '22}
}

@inproceedings{xgboost_model, series={KDD ’16},
   title={XGBoost: A Scalable Tree Boosting System},
   url={http://dx.doi.org/10.1145/2939672.2939785},
   DOI={10.1145/2939672.2939785},
   booktitle={Proceedings of the 22nd ACM SIGKDD International Conference on Knowledge Discovery and Data Mining},
   publisher={ACM},
   author={Chen, Tianqi and Guestrin, Carlos},
   year={2016},
   month=aug, pages={785–794},
   collection={KDD ’16} }

@article{1-nn,
author = {Cover, T. and Hart, P.},
title = {Nearest neighbor pattern classification},
year = {2006},
issue_date = {January 1967},
publisher = {IEEE Press},
volume = {13},
number = {1},
issn = {0018-9448},
url = {https://doi.org/10.1109/TIT.1967.1053964},
doi = {10.1109/TIT.1967.1053964},
journal = {IEEE Trans. Inf. Theor.},
month = sep,
pages = {21–27},
numpages = {7}
}

@inproceedings {website_fingerpriting,
author = {Tao Wang and Xiang Cai and Rishab Nithyanand and Rob Johnson and Ian Goldberg},
title = {Effective Attacks and Provable Defenses for Website Fingerprinting},
booktitle = {23rd USENIX Security Symposium (USENIX Security 14)},
year = {2014},
isbn = {978-1-931971-15-7},
address = {San Diego, CA},
pages = {143--157},
url = {https://www.usenix.org/conference/usenixsecurity14/technical-sessions/presentation/wang_tao},
publisher = {USENIX Association},
month = aug
}

@ARTICLE{intrusion_detection,
  author={Kumar, Satish and Gupta, Sunanda and Arora, Sakshi},
  journal={IEEE Access}, 
  title={Research Trends in Network-Based Intrusion Detection Systems: A Review}, 
  year={2021},
  volume={9},
  number={},
  pages={157761-157779},
  doi={10.1109/ACCESS.2021.3129775}}

\end{document}